# Optical phonon modes of wurtzite InP

E. G. Gadret[1,2], T. Chiaramonte[1], M. A. Cotta[1], F. Iikawa[1,a)], M. M. de Lima, Jr.[2], A. Cantarero[2], J. R. Madureira[2,3]

[1]*Instituto de Física "GlebWataghin", Unicamp, 13083-859, Campinas, Brazil*
[2]*Materials Science Institute, University of Valencia, P.O. Box 22085, E-46071 Valencia, Spain*
[3]*Faculdade de Ciências Integradas do Pontal, Universidade Federal de Uberlândia (UFU), 38304-402, Ituiutaba, Brazil*

Optical vibration modes of InP nanowires in the wurtzite phase were investigated by Raman scattering spectroscopy. The wires were grown along the [0001] axis by the vapor-liquid-solid method. The $A_1$(TO), $E_{2h}$ and $E_1$(TO) phonon modes of the wurtzite symmetry were identified by using light linearly polarized along different directions in backscattering configuration. Additionally, forbidden longitudinal optical modes have also been observed. Furthermore, by applying an extended 11-parameter rigid-ion model the complete dispersion relations of InP in the wurtzite phase have been calculated, showing a good agreement with the Raman experimental data.

Semiconductor nanostructures based on III-V compounds have been extensively investigated in the last decades due to the high quality of these materials and their outstanding optical and electronic properties. As a consequence, they have been employed in optical device applications, as well as for fundamental physics investigation [1-6]. In general, bulk III-V compounds can crystallize in both hexagonal and cubic crystal structures, depending on the specific elements involved. In particular, phosphide and arsenide III-V compounds typically grow in the cubic phase, which for these materials is the most stable at ambient pressure. As a consequence, they have been largely investigated in the last few decades. More recently, the advent of semiconductors nanowires (NWs) has allowed the growth of III-V compounds in crystal phases that are different from the bulk, such as the hexagonal wurtzite (WZ) phase [7-10] for both phosphides and arsenides.

In contrast to the optical properties of III-V WZ phase that are relatively well-known [4-7], its vibration properties are not deeply known. In nanostructured devices the thermal dissipation is a crucial point for device application. For instance, the thermal conductivity is generally dominated by the phonon modes for undoped large and middle band gap semiconductors [11]. Furthermore, due to the scarce available experimental and theoretical data of III-V compounds WZ phase, as well as their phonon dispersion, so far few works have been investigated their thermal properties.

Among the III-V compound NWs, InP is one of the best known materials, because of the interest raised by their relatively efficient optical emission due to the low carrier capture velocity by surface states in In-based compounds [12], which is one order of magnitude smaller than that for GaAs. InP NWs also present ZB or WZ phases depending on the growth conditions and, in some cases, both phases coexist when the wire axis is oriented along the [0001] direction of WZ or the [111] of ZB phase. The coexistence of two phases significantly affects the optical properties due to the existence of a band offset between ZB and WZ band structures. Concerning the vibration modes of WZ InP, extra phonon modes appear at the center of the Brillouin zone as compared to those for ZB phase. These additional phonon modes can be represented as a fold of Γ-L path of the Brillouin zone for ZB phase along [111]-direction to Γ-A dispersion for WZ phase along [0001]-axis, as shown in Figure 1.

The WZ phonon modes for III-V compounds have been recently investigated in InN, GaN, InAs, GaAs and InP by Raman scattering [13-17]. However, as for InP, to the best of our knowledge, a detailed investigation of the Raman modes in WZ InP NWs has not been reported so far. Surprisingly, the results reported by refs. [18] and [19] present different attribution for the Raman peaks observed in WZ InP. Therefore, in this paper, we performed detailed theoretical and experimental investigation on the vibration properties of WZ InP NWs.

The selection rules for the Raman scattering of a WZ crystal are well known [20]. The Raman intensity is obtained from the well know expression [21]:

$$I \propto |\hat{e}_s . R_k . \hat{e}_i|^2, \qquad \text{Eq. 1}$$

Where $\hat{e}_i$ and $\hat{e}_s$ are polarization vectors of incident and scattered light, respectively, and $R_k$ is the Raman tensor corresponding to the $k$ phonon mode. In Table I, we present the allowed phonon modes in backscattering configuration used in this work. The light propagation direction in our case is along the $x$-axis of the crystal, and light polarizations along $y$ or $z$-axis. The allowed optical modes for different polarization configurations using Porto´s notation [22] are also shown in the Table I. In our experimental configuration, the allowed Raman modes are only transversal-optical (TO) modes, while all longitudinal-optical (LO) modes are forbidden.

---

[a] Author to whom correspondence should be addressed. Electronic mail: iikawa@ifi.unicamp.br

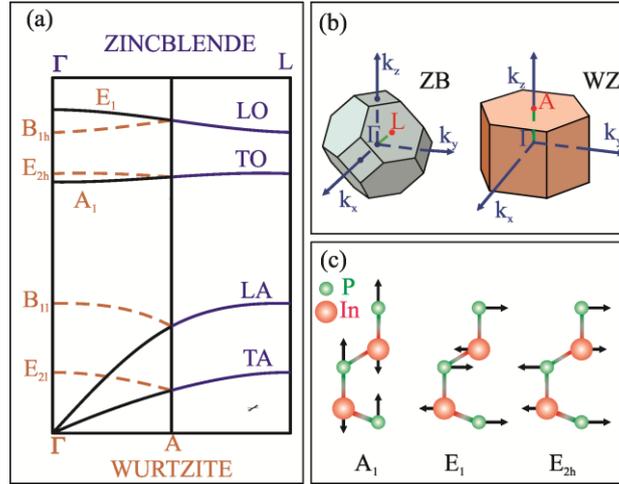

Figure 1 – *(a) Diagram of phonon folding from the ZB (along [111]) to the WZ (along [0001]) phase; (b) Brillouin zones for ZB and WZ crystal phases;(c) atomic displacement for the WZ vibration modes.*

In order to calculate the phonon dispersion for WZ InP, we used an extended 11-parameter rigid-ion model [23]. In this way we can calculate the phonon dispersion relation energy for WZ structure by parameterizing the already known ZB structure dispersion relation. In our calculation we used the experimental dispersion relation of the ZB InP from Refs. [24] and [25]. From the evaluated fitting parameters above, we calculated WZ InP dispersion relation frequency which is shown in Fig. 2. The values of the optical phonon mode frequencies at Γ-point are presented in Table I.

InP nanowires were grown by the vapor-liquid-solid (VLS) method in a Chemical Beam Epitaxy (CBE) system using ~25 nm size Au nanoparticles as catalyst. The nanowires were mechanically transferred to a GaAs wafer, which contains a 5 μm thick Al film. The Al layer is used to avoid the spurious optical signal from the GaAs, to increase the optical contrast between NWs and substrate, which is helpful for micro-Raman measurements and to help in the power dissipation. We investigated stacking fault free pure WZ phase InP nanowires, which were grown at 420°C using a Trimethil-indium (TMIn) flux of 0.15 standard cubic centimeters per minute. In a previous work, the transmission electron microscope results for the same sample show only pure WZ phase. All NWs are grown along the [0001] axis, with 5-10 μm length and 50-100 nm diameter [6].

For Raman measurements we used the 488 nm line of an Ar$^+$ laser as excitation, a T64000 Jobin-Yvon spectrometer, coupled to an open electrode liquid N$_2$ cooled Si-CCD, and an olympus optical microscope. We used 100× optical objective (numerical aperture 0.95) to both focus the laser beam and collect the scattered light. The sample holder system contains motorized XYZ positioner. In order to follow the Raman selection rule discussed above, we used a λ/2-plate, in order to rotate the incident linear polarization of the laser beam, and a linear analyzer for the scattered light placed before the monochromator entrance in order to keep the light polarization along the largest response of the monochromator diffraction grating. All experiments shown in this work were performed at room temperature.

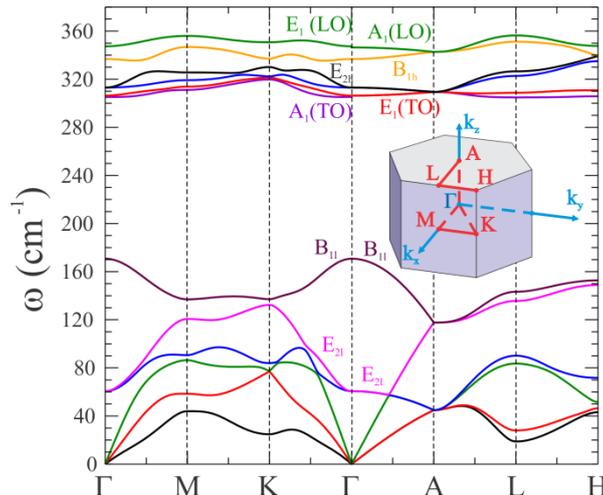

Figure 2 - *Calculated phonon dispersion relations of InP in the WZ phase.*

Figure 3 shows a schematic diagram of the sample geometry used in our experimental setup. The X, Y, and Z are the laboratory coordinates. The NW-axis is along [0001], parallel to Z-axis in Fig. 3, the laser beam is along X-axis, and scattered light direction is in the opposite direction (backscattering configuration). The NWs are thus perpendicular to the direction of the incident and scattered light. The incident light polarization can be rotated in the YZ-plane. The NWs measured here are in directions along Y or Z axis.

In order to compare the Raman selection rules with our experimental setup we have to transform the laboratory coordinates to the usual sample crystallographic ones (*x*, *y*, and *z*). The unique crystallographic axis that we have obtained is [0001], which is along the wire direction. For a given crystal plane, the incident and scattered light directions and their polarizations defines the Raman scattering selection rules. The crystal plane of the lateral side of the NWs, in our case, is unknown. Fortunately, our experimental condition is that the selection rule does not change if we rotate the NW in the *z*-axis by an angle θ. The Raman intensity given by Eq. 1 follows the same selection rules shown in Table I for any angle θ between the *x*-axis, perpendicular to the wire, and X-axis of the laboratory, considering *z* parallel to the Z-axis.

Table I – *Optical phonon modes and polarization configurations* in *backscattering geometry for WZ InP.*

| Phonon Modes | Polarization Configurations | Exp. ω (cm$^{-1}$) | Calc. ω (cm$^{-1}$) |
|---|---|---|---|
| $A_1$(TO) | $x(zz)\bar{x}$ ; $x(yy)\bar{x}$ | (302.1±0.8) | 305.3 |
| $E_1$(TO) | $x(zy)\bar{x}$ ; $x(yz)\bar{x}$ | (302.4±0.8) | 306.3 |
| $E_{2h}$ | $x(yy)\bar{x}$ | (306.4±0.7) | 313.0 |
| $A_1$(LO) | forbidden | (341.9±0.8) | 346.4 |
| $E_1$(LO) |  | not observed | 347.3 |

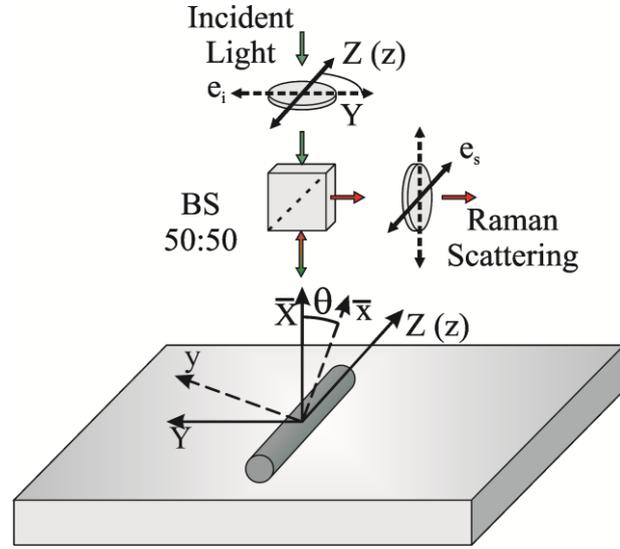

Figure 3 - *Schematic diagram of the polarization configurations used for Raman scattering, where X, Y and Z are the laboratory coordinates and x, y and z are the sample coordinates.*

We used four different polarization configurations: $x(yy)\bar{x}$, $x(yz)\bar{x}$, $x(zy)\bar{x}$, and $x(zz)\bar{x}$, the same configurations shown in the Table I. Figure 4 depicts typical Raman spectra measured in a single WZ-InP NW for four distinct polarization configurations. We analyzed approximately eight single NWs and in all of them the Raman spectra are very similar to those shown in Fig. 4. The spectra are corrected by the optical response of all optical components. All spectra are well fitted using Lorentzian functions.

The Raman spectra of Fig. 4 show the main Raman lines in the *300 - 310* cm$^{-1}$ range in all polarization configurations, around the TO phonon modes. We observe remarkably stronger Raman lines in parallel configurations, $x(yy)\bar{x}$ and $x(zz)\bar{x}$, than in crossed polarizations, $x(yz)\bar{x}$ and $x(zy)\bar{x}$. In Table I, $x(zz)\bar{x}$ polarization selects the $A_1$(TO) mode, while, $x(zz)\bar{x}$ selects both the $A_1$(TO) and $E_{2h}$ modes. Therefore, the strongest peak at 302.1 cm$^{-1}$ in the spectrum for $x(zz)\bar{x}$ polarization is attributed to the $A_1$(TO) vibration mode. The Raman spectrum for $x(yy)\bar{x}$ polarization also shows two peaks, a strong one at 306.4 cm$^{-1}$ and a weak shoulder at 302.1 cm$^{-1}$. The latter presents practically the same position of the peak observed in $x(zz)\bar{x}$ configuration, attributed to the $A_1$(TO) mode, and the former, by exclusion, is attributed to the $E_{2h}$ mode.

The slight lower Raman intensity for $x(yy)\bar{x}$ polarization as compared to that of $x(zz)\bar{x}$ is attributed to the dielectric contrast effect [26,27]. This effect occurs for cylindrical shaped NWs when the diameter is smaller than the wavelength of the light and if the dielectric constant of the NW is very different from that of the environment (air or vacuum) [6,7]. The intensity of the transmitted light into the wire, which is polarized perpendicular to the wire, is reduced, while the light polarized along the wire remains unchanged.

The Raman spectra for cross polarization configuration, $x(yz)\bar{x}$ and $x(zy)\bar{x}$, are relatively similar to each other as expected, and present a peak at 302.4 cm$^{-1}$ and 307.1 cm$^{-1}$. Based on the selection rules shown in Table I, only the $E_1$(TO) mode is allowed. This mode is usually close to the $A_1$(TO) mode, as observed in our theoretical results (see Table I and Fig. 2), therefore, the peak at 302.4 cm$^{-1}$ can be attributed to the $E_1$(TO). The other peak at 307.1 cm$^{-1}$ can be assigned as the $E_{2h}$ mode, because it is close to the value obtained for this mode in parallel configurations and it is observed in all polarization configurations. Note that the frequency difference between the observed $A_1$(TO) and $E_1$(TO) modes are, within the experimental error, in agreement with the calculated values.

In a previous work, Lohn *et al* report [18] on the Raman scattering results in an ensemble of InP NWs, containing both ZB and WZ phases. The observed Raman spectra present peak positions very similar to ours. Since the WZ peaks are overlapped with that of the ZB in their spectrum, the broad peak at the TO-mode range is attributed to the superposition of $E_1(TO)$, $A_1(TO)$, $E_{2h}$ and $F_2$ (ZB mode) phonon modes. Furthermore, they could not use the different polarization configuration, since the NWs were randomly oriented. On a different work, Chashnikova *et al.* [19], using only crossed polarization configurations, identified the $E_1(TO)$ mode at a similar frequency of that obtained in our work at the $E_1(TO)$ position (see Figure 4), and the $E_{2h}$, in contrast to our theoretical and experimental result, at a lower frequency range.

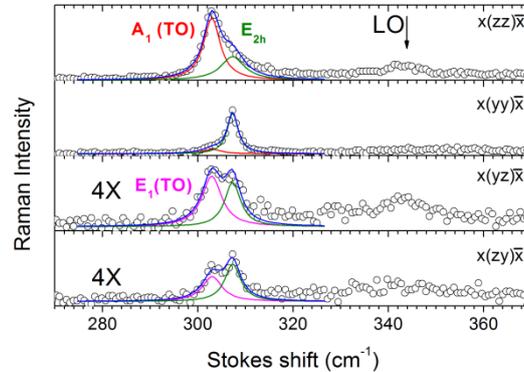

Figure 4 - *Raman spectra from single WZ-InP NW at four different polarization configurations.*

The Raman spectra of Fig. 4 show an additional weak peak at 341.9 cm$^{-1}$, which is in the spectral range of the forbidden LO-phonon modes, $E_1(LO)$ and $A_1(LO)$. For backscattering configuration only $A_1(LO)$ mode is allowed, but when the incident light is along the [0001] axis, while $E_1(LO)$ mode is always forbidden in backscattering configuration. The presence of this mode could be related to a resonance effect, since the laser line used here (488 nm – 2.54 eV) can be close to the $E_1$ electronic transition in ZB crystals. This transition is around 3.0 eV at 80 K for InP [28], however, the exact value of the equivalent transition is unknown for the WZ phase. In addition, the LO-mode has also been observed in other III-V compounds NWs, such as GaAs [13] and InAs [14] NWs, in Raman scattering using similar experimental condition and have been attributed to a quasi-resonance effect. In summary, the calculated TO and LO-modes for WZ phase are in good agreement with our experimental data shown in Table I.

**Conclusions**

In this work, we present Raman scattering results of the optical vibration modes of InP NWs in the wurtzite phase. These results are supported by theoretical calculations performed by means of an 11-parameters rigid ion model. Using different polarization configurations we have identified three optical phonon modes, $A_1(TO)$, $E_{2h}$ and $E_1(TO)$. We also observed the selection-rule-forbidden LO-phonon mode. The calculated vibration modes are in good agreement with the experimental results. These results can be helpful for investigating the thermal properties and also the electron-phonon interaction in WZ phase InP.

**Acknowledgment**
The authors acknowledge CNPq, FAPESP, the Spanish Ministry of Science and Innovation (grant MAT2009-10350), and bilateral CAPES Brazil/MICINN Spain cooperation agreement for financial support. We also acknowledge Dr. M. Möller for discussion and laboratory assistance.